\documentclass[11pt,letterpaper]{article}
\usepackage[utf8]{inputenc}
\usepackage[margin=1in]{geometry}
\usepackage{graphicx}
\usepackage{booktabs}
\usepackage{amsmath,amssymb}
\usepackage[labelfont=bf,font=small]{caption}
\usepackage{float}
\usepackage[numbers,sort&compress]{natbib}
\usepackage{url}
\usepackage[colorlinks=true,allcolors=blue]{hyperref}
\title{Image-Based Breast Implant Detection for Mammography Dataset Curation and Near-Real-Time Deployment: Comparing Foundation Models and Task-Specific Convolutional Models}
\author{\parbox{\textwidth}{\centering
Vasisht Ishwar$^{1,2}$, Hari Trivedi, MD$^{1}$, Young Seok Jeon, PhD$^{1}$,
Beatrice Brown-Mulry, MS$^{1}$, Frank Li, PhD$^{1}$, Rohan Satya Isaac, MS$^{1}$,
Mohammadreza Chavoshi, MD$^{1}$, Judy Wawira Gichoya, MD, MBChB, MS$^{1}$\\[1em]
$^{1}$Department of Radiology and Imaging Sciences, Emory University School of Medicine, Atlanta, GA, USA\\
$^{2}$University of Maryland, College Park, MD, USA\\[1em]
\normalsize Corresponding author: Judy Wawira Gichoya, MD (\texttt{judywawira@emory.edu})
}}
\date{}

\begin{document}
\maketitle

\begin{abstract}
\textbf{Purpose:}

To evaluate the performance--feasibility tradeoffs of foundation models (FMs) and task-specific convolutional neural networks (CNNs) trained from scratch for breast implant classification in 2D mammography, with emphasis on suitability for near real-time clinical deployment.

\textbf{Methods:}

We evaluated four models: two FMs (RAD-DINO and MammoCLIP) and two CNNs trained from scratch for implant prediction (ResNet18 and our lightweight ResNetLite). Using the Emory Breast Imaging Dataset, 5,000 unilateral screening mammograms were used for training/validation and 1,000 manually reviewed unilateral images were held out for testing. For the FMs, global image embeddings from the pretrained encoder were classified using a support vector machine (SVM). The CNNs were trained end-to-end on 2D mammograms, with ResNetLite optimized via grid search over depth and width to balance accuracy and efficiency. Performance was evaluated using AUROC, sensitivity, specificity, accuracy, embedding visualization, and inference-latency.

\textbf{Results:}

All models demonstrated strong performance on held-out test data ($n$ = 1,000). MammoCLIP achieved the highest AUROC (0.999) with the quickest training time of 493 seconds. RAD-DINO achieved the highest sensitivity (0.980; accuracy 0.989) but had the slowest inference and training times. ResNet18 and MammoCLIP achieved comparable accuracy (0.985). ResNetLite showed no statistically significant difference from ResNet18 (AUROC 0.993; accuracy 0.976) despite using only 1.4\% of ResNet18's parameters, and had the fastest inference time.

\textbf{Conclusion:}

FMs and task-specific CNN models reliably detect breast implants on 2D mammography. Model selection is best guided by deployment context: MammoCLIP for GPU-equipped hospital settings requiring scalable integration, and lightweight CNNs such as ResNetLite for resource-constrained or edge deployments.
\end{abstract}


\section{Introduction}

Screening mammography is recommended annually in the United States for women beginning at age 40 to enable early breast cancer detection. Many women who undergo screening have breast implants for cosmetic or reconstructive reasons, which distort breast tissue appearance, obscure parenchymal regions and can reduce mammographic sensitivity by up to 30\%~\cite{r1,r2,r3}. Breast implant prevalence is estimated at roughly 2--3\% in population-based European screening cohorts~\cite{r4}, while U.S. screening populations tend to run higher and more variable, given greater augmentation rates and regional concentration in the Pacific/Mountain West; with over 300,000 augmentation procedures performed annually in the U.S. alone~\cite{r5} and most patients augmented in their 20s and 30s now aging into screening-eligible ranges, this prevalence is expected to keep climbing.

While many research and commercial models have now been developed~\cite{r6,r7} for breast cancer detection, these models do not report performance in patients with breast implants and the 510k clearance~\cite{r8} for these models do not state any such requirement. Such distribution shifts have been shown to substantially negatively affect model performance~\cite{r9}, resulting in a scenario in which an AI model may be generating cancer predictions in clinical deployment that are significantly worse than non-implant cases. Such models may benefit from exclusion of implant containing-cases at deployment, or development of specialized models for patients with breast implants. Detecting the presence of implants in mammography images is therefore both an important research, development, and translational task for AI in breast imaging~\cite{r10,r11}.

To our knowledge, there are currently no publicly available models specifically to detect implants in mammograms, in part due to limited labeled datasets and inconsistent implant documentation in radiology reports and DICOM metadata~\cite{r4,r12,r13}. A common method of detecting the presence of implants is the use of DICOM metadata or associated clinical documentation, both of which may contain omissions or inaccuracies~\cite{r14,r15}. Such inconsistencies can propagate errors into curated datasets and downstream artificial intelligence (AI) workflows, reducing the reliability of trained models~\cite{r9,r16}. Our previous work has shown models can be significantly biased by taking `shortcuts' from confounding imaging features \cite{r17}.

While a model for implant detection could be trained end-to-end using standard CNN techniques~\cite{r18}, recent advances in medical imaging foundation models (FMs) have demonstrated strong performance across a wide range of radiologic tasks through large-scale pretraining and transferable feature representations~\cite{r19,r20,r21,r22,r23}. FMs may be particularly advantageous for implant detection because they leverage large-scale pretraining to learn transferable radiologic feature representations without requiring task-specific architecture development or extensive retraining~\cite{r21,r22,r24,r25}. Given the variability in mammographic acquisition protocols across institutions, pretrained FMs may provide a scalable approach capable of adapting to diverse imaging environments while minimizing the need for manually engineered solutions. However, the advantages of these large models for relatively constrained binary classification tasks such as implant detection remain unclear~\cite{r19,r20,r21,r22,r23,r24,r25}. Because mammography systems are frequently deployed across environments with varying computational resources, understanding the tradeoff between model complexity and performance is important for scalable real-world implementation~\cite{r9,r16}.

In this study, we evaluate the performance, computational efficiency, and cost of two pretrained FMs, RAD-DINO~\cite{r26} and MammoCLIP~\cite{r27}, alongside two convolutional neural networks, ResNet18~\cite{r18} and a custom lightweight architecture, ResNetLite, for breast implant detection on screening mammography.

\section{Methods}

\subsection{Study Design and Data Source}

In this retrospective study, we utilized 6,000 randomly sampled 2D screening mammographic images from the Emory Breast Imaging Dataset (EMBED)~\cite{r28}, which contains 1.8M 2D images over a 10 year period with varying sites and acquisition protocols. The EMBED dataset contains two main components: (1) DICOM metadata at the image level, and (2) Clinical data, which includes information entered by radiologists and technologists such as exam type, imaging findings, and outcomes.

\begin{figure}[tbp]
\centering
\includegraphics[width=0.64\textwidth]{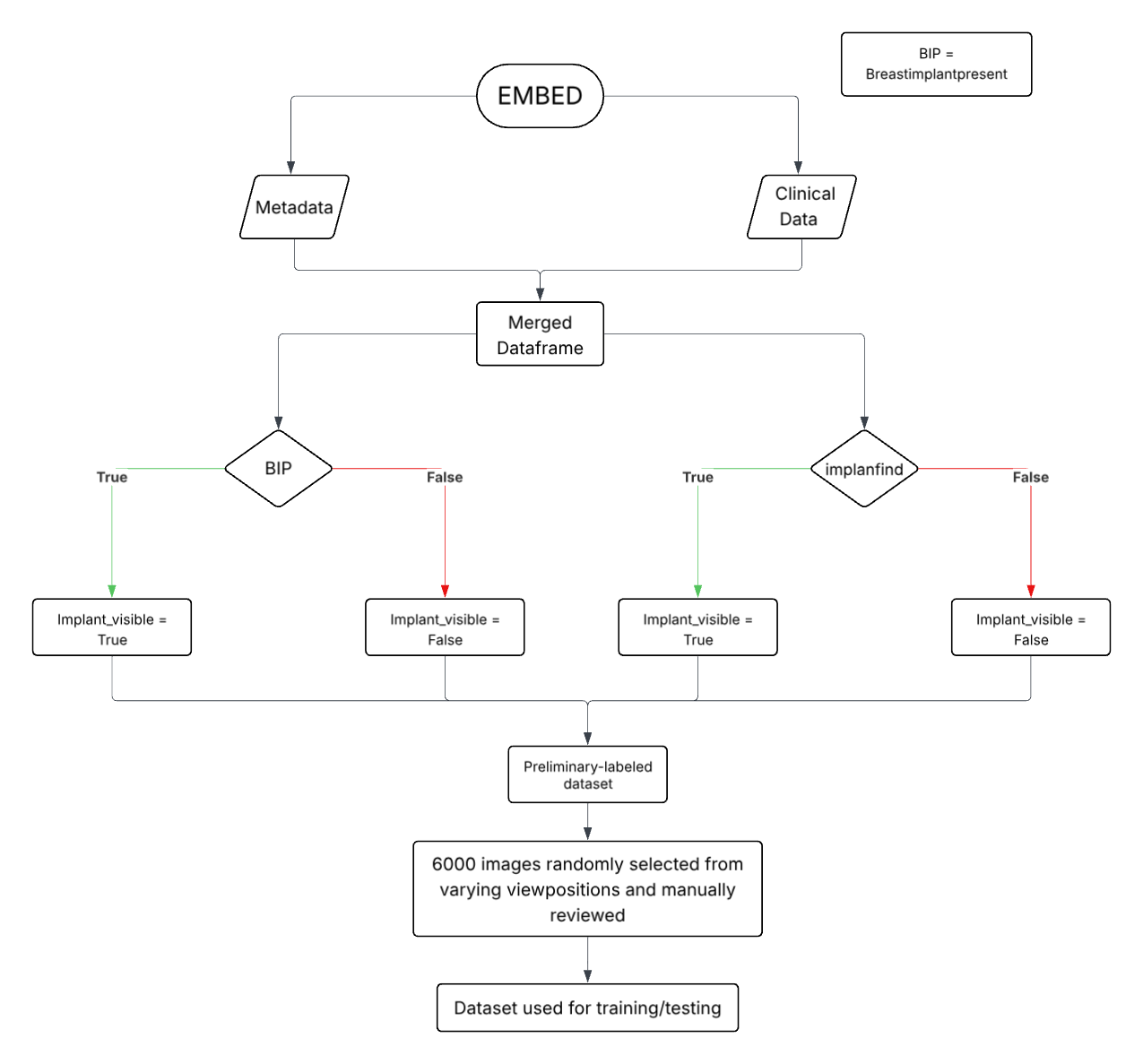}
\caption{\textbf{Data curation workflow for the implant visibility dataset.} The process starts with the EMBED dataset, merging the metadata and clinical data to establish ground truth labels. Images were preliminarily based on ``BreastImplantPresent'' (BIP) status and ``implanfind'' tags that could be found in the metadata and clinical data respectively. A subset of 6,000 images, balanced across varying view positions, underwent manual review to generate the final dataset used for training and testing.}
\label{fig:1}
\end{figure}

Preliminary image-level implant labels were derived from both DICOM metadata and Clinical data. We used the \emph{BreastImplantPresent} tag from DICOM metadata and \emph{implanfind} field populated by a radiologist from the Clinical data. If either was present, the image was preliminary categorized as ``implant\_visible'' (Fig.~\ref{fig:1}). In addition, we used the \emph{ViewPosition} tag to identify exams with implants by searching for Implant Displaced (ID) views, and using the correspond non-implant displaced view. For example, an exam with a mediolateral oblique implant displaced (MLOID) view could be assumed to contain a visible implant in its MLO view. All images were then manually reviewed and relabeled appropriately as positive or negative until a target of 3,000 images per class was achieved. Both implant-containing and implant-displaced mammography views were included, as the displaced views frequently did not exclude the entire implant~\cite{r4}.

\subsection{Image Preprocessing}

Original DICOM files were read using PyDICOM and, when present, had VOI lookup tables applied to correctly map stored pixel values to display values~\cite{r29}. Each image was rescaled to have pixel intensities in the range $[-1,1]$. Images were converted to RGB format to match the input requirements of ResNets. Finally, all images were resized to $224\times224$ pixels.

\subsection{Dataset Partitioning}

\begin{figure}[tbp]
\centering
\includegraphics[width=0.7\textwidth]{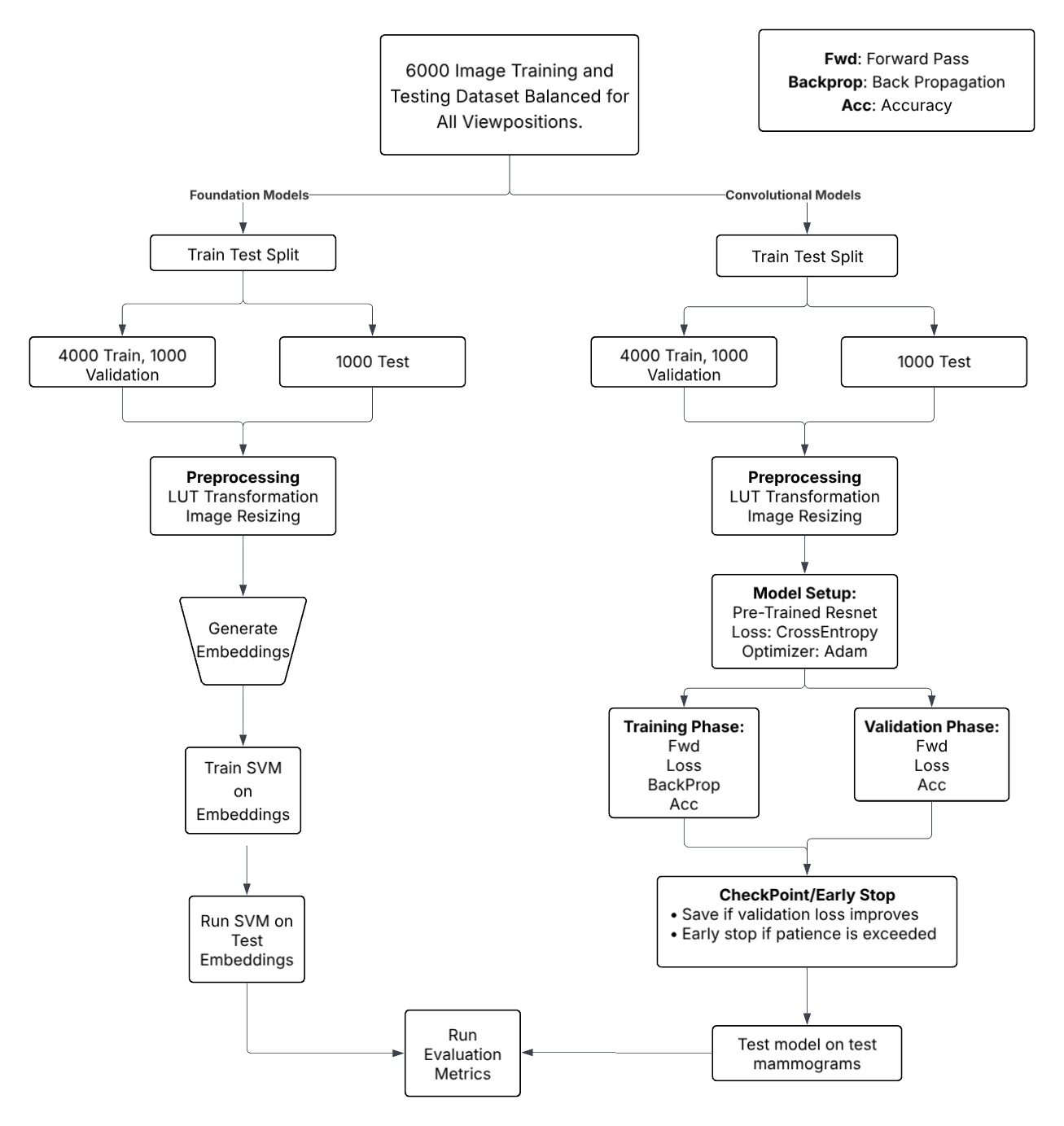}
\caption{\textbf{Overview of the dual modeling pipeline using the balanced 6,000-image dataset.} The workflow diverges into two approaches: (left) A FM approach utilizing feature extraction and support vector machine (SVM) classification, and (right) a convolutional neural network (CNN) approach using a pre-trained ResNet. Both pipelines utilize identical train/validation/test splits (4,000/1,000/1,000) and preprocessing steps (LUT transformation and resizing) before converging on final evaluation metrics.}
\label{fig:2}
\end{figure}

From our dataset, we split a total of 6,000 image-label pairs into three subsets: training (66.7\%), validation (16.7\%), and test (16.7\%). Partitioning was performed at the patient level to avoid data leakage across subsets. We used stratified randomization based on implant status to keep the class distribution balanced across training, validation, and test sets (Fig.~\ref{fig:2}).

\subsection{Foundation Models}

We explored two complementary approaches to feature extraction using state of the art pretrained FMs. We first used MammoCLIP, a multimodal model specifically trained on large-scale mammography datasets paired with associated textual reports. MammoCLIP's vision encoder generates 512-dimensional embeddings that capture fine-grained features unique to breast imaging, including differences across views and imaging systems. MammoCLIP's training only on mammograms makes it well-suited for tasks requiring detailed breast-tissue representation. Nevertheless, its performance may be sensitive to the distribution of cases present in its training data, which could limit its ability to generalize to underrepresented or challenging cases such as patients with dense tissue or ruptured implants.

The second complementary approach was implemented using RAD-DINO, a vision transformer model trained exclusively on chest radiographs. RAD-DINO outputs 768-dimensional embeddings and has been shown to learn robust radiographic representations that are transferable to other projection imaging tasks~\cite{r26}. Although its training set did not include mammograms, the shared imaging characteristics between chest X-rays and mammography, including contrast dynamics, anatomical structures with soft-tissue/bone interfaces, and the projection-based nature of acquisition, suggest that the learned representations can still be informative for downstream mammographic classification tasks. By evaluating both MammoCLIP and RAD-DINO, our goal was to compare the benefits of highly domain-specific mammography representations against more broadly generalizable radiography representations.

\subsection{Convolutional Models}

\begin{figure}[tbp]
\centering
\makebox[\textwidth][c]{\includegraphics[width=1.2\textwidth]{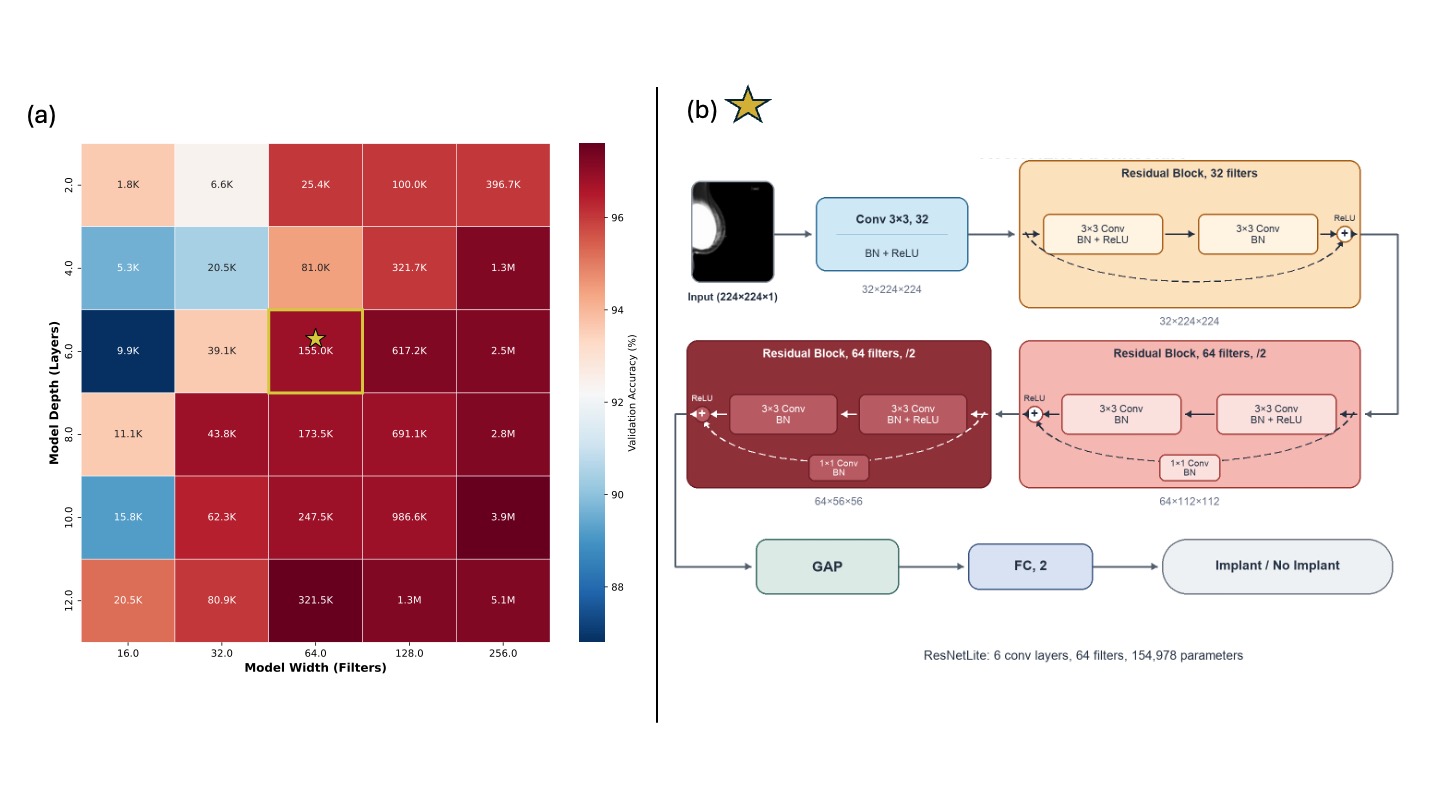}}
\caption{\textbf{ResNetLite optimization and final architecture for implant detection.} (a) Grid search heatmap displaying the trade-off between model depth and width against validation accuracy. Cell contents specify the total parameters. The highlighted cell (star) indicates the optimal configuration (155.0K parameters) selected for efficiency. (b) The network takes a $224\times224\times1$ grayscale mammography image as input and applies a $3\times3$ stem convolution (32 channels), followed by three residual stages. Stage 1 (32 filters, stride 1) preserves spatial resolution with an identity skip connection. Stages 2 and 3 (64 filters each, stride 2) each halve the spatial resolution ($224\times224 \rightarrow 112\times112 \rightarrow 56\times56$) and use a $1\times1$ convolution on the skip connection to match dimensions for the residual addition. Each residual block contains two $3\times3$ convolutional layers, for a total of 6 convolutional layers across the network. A global average pooling (GAP) layer and a fully connected (FC) layer produce the final binary classification (implant / no implant).}
\label{fig:3}
\end{figure}

We implemented two CNNs for comparison: ResNet18 and our custom architecture, which we termed ResNetLite. ResNetLite is a custom lightweight variant inspired by the ResNet18 architecture, designed to maximize computational efficiency and inference speed while maintaining competitive performance. To optimize ResNetLite, we performed a grid search over architectural depth and width, systematically reducing the number of convolutional layers and filters until we achieved a balance between accuracy and computational cost. This search revealed that ResNetLite achieved its optimal performance at a depth of six layers and a maximum width of 64 channels, which we used as the final configuration in all experiments. (Fig.~\ref{fig:3})

Both ResNet18 and ResNetLite were trained end-to-end on 2D mammogram images, initialized with ImageNet-pretrained weights and fine-tuned on our curated mammography dataset using binary cross-entropy loss for implant detection (Fig.~\ref{fig:2}). To mitigate the impact of class imbalance, class weights were assigned inversely proportional to the prevalence of each class in the training set. Optimization was performed using Adam~\cite{r30} optimizer with an initial learning rate of $10^{-4}$ and a batch size of 32. Early stopping with a patience of 10 epochs based on validation loss was employed to prevent overfitting and ensure generalization (Fig.~\ref{fig:2}).

\subsection{Model Evaluation}

For both FMs, the global image embeddings were classified using a support vector machine (SVM) to evaluate their predictive performance. For the CNNs, model predictions were generated by applying a softmax function to the output logits to obtain class probabilities, with the predicted class assigned via argmax (equivalent to a 0.5 probability threshold).

Performance for all models was assessed using sensitivity (recall), specificity, precision, F1-scores, overall accuracy, inference latencies, area under the receiver operating characteristic curve (AUROC), and area under the precision-recall curve (AUPRC). In addition to quantitative evaluation, we inspected the learned representation spaces (embeddings) for each model. For FMs, we extracted embeddings from the final encoder layer without additional fine-tuning and trained a lightweight classification head on top of the embeddings. For CNNs, we extracted embeddings using a forward hook on the final average pooling layer, capturing high-level feature representations immediately prior to classification. Pairwise differences in model performance were tested for statistical significance using McNemar's exact test and 95\% confidence intervals. 95\% confidence intervals for each model's individual metrics were generated via percentile bootstrap resampling of the test set (2,000 resamples), while statistically significant differences between specific model pairs were assessed using paired bootstrap resampling of the same resampled test images (2,000 resamples). All models were evaluated on an NVIDIA L40S GPU.

\section{Results}

\subsection{Dataset Characteristics}

Our final dataset consisted of 3,000 positive and 3,000 negative images. Of note, the negative cases are a combination of patients with no implants and ID views in which the implant is \emph{fully} displaced and therefore not visible. The positive cases contain non-implant displaced views and ID views in which the implant could only be partially displaced. One hundred and twelve images (1.9\%) in the positive set contained an implant that was only partially displaced. View positions were roughly balanced: 3,064 images (51.1\%) were acquired in the MLO view, 2,824 images (47.0\%) in the CC view (Table~\ref{tab:1}).

\begin{table}[tbp]
\centering
\caption{Composition of the final 6,000-image dataset, showing the balanced split between implant-visible and non-visible cases and the distribution across mammographic view positions.}
\label{tab:1}
\begin{tabular}{llrr}
\toprule
\textbf{Category} & \textbf{Subcategory} & \textbf{Number of Images} & \textbf{Percentage of Total} \\
\midrule
\textbf{Total Images} & & 6000 & 100\% \\
\midrule
\textbf{Implant Visibility} & Visible (True) & 3000 & 50\% \\
 & Not Visible (False) & 3000 & 50\% \\
\midrule
\textbf{View Position} & MLO (Mediolateral Oblique) & 3064 & 51.1\% \\
 & CC (Craniocaudal) & 2824 & 47.0\% \\
 & Implant-Displaced & 112 & 1.9\% \\
\bottomrule
\end{tabular}
\end{table}

\begin{figure}[tbp]
\centering
\includegraphics[width=0.58\textwidth]{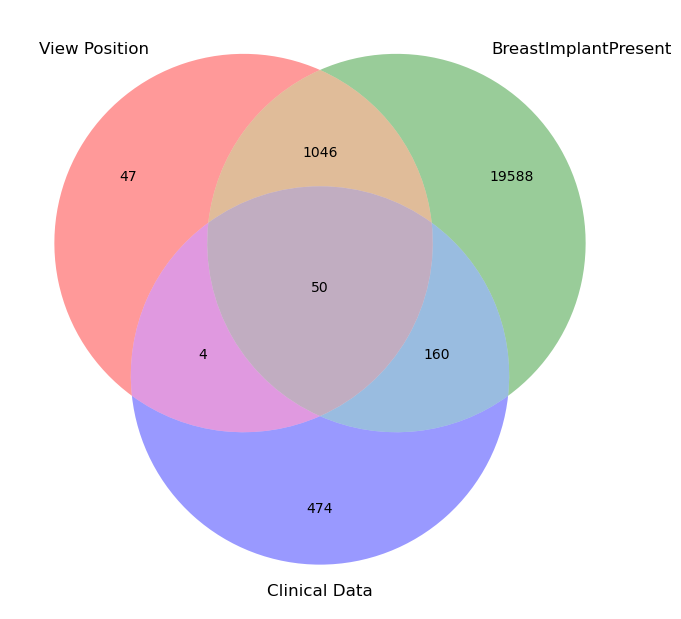}
\caption{\textbf{Overlap of implant-containing exams across label sources.} Venn diagram illustrating the overlap of exams of patients containing implants between three sources of labels to identify implant visibility: DICOM ``BreastImplantPresent'' tags, Clinical Data ``implanfind'' tags, and View Position tags. The minimal intersection of all three sources from the entire EMBED dataset ($n$ = 50) compared to the large unique subset found in the DICOM data ($n$ = 19588) demonstrates poor agreement between the tags, indicating that each source captures a partially distinct subset of implant cases.}
\label{fig:4}
\end{figure}

Approximately 23\% of images were initially mislabeled using the DICOM and Clinical data and were subsequently corrected by the annotator prior to model testing and validation. The Venn diagram in Figure~\ref{fig:4} illustrates the inconsistency among metadata sources used to identify exams where implant cases were present, highlighting the unreliability of relying solely on data curation methods for feature extraction and the necessity for manual review. Specifically, poor overlap between the DICOM (BreastImplantPresent tag), Clinical Data (implanfind tag), and ViewPosition tags indicates that each source captured a distinct subset of implant cases, with limited agreement across systems.

\subsection{Comparison of Foundation Models}

RAD-DINO and MammoCLIP demonstrated strong overall performance, with results approaching near-perfect discrimination (Table~\ref{tab:2}). RAD-DINO achieved slightly higher sensitivity (0.9802 vs. 0.9722) than MammoCLIP; however, their 95\% confidence intervals overlapped substantially, indicating this difference is not statistically significant, consistent with the McNemar comparison ($p$ = 0.125). Both models achieved identical specificity (0.9980). RAD-DINO showed higher precision and F1-scores than MammoCLIP, though the differences were small and their confidence intervals also overlapped (Table~\ref{tab:2}), indicating these differences are not statistically robust either.

\begin{table}[tbp]
\centering
\caption{Standard evaluation metrics for: (1) a lightweight ResNet (ResNetLite) model, (2) a ResNet18 convolutional neural network, (3) a support vector machine (SVM) classifier trained on embeddings from the MammoCLIP FM, and (4) a SVM trained on embeddings from the RAD-DINO FM. Metrics include sensitivity, specificity, precision, F1-score, and accuracy, along with AUROC and AUPRC. A row for model size (Parameters) is included to highlight the large efficiency gains of ResNetLite relative to MammoCLIP, RAD-DINO, and ResNet18, alongside rows for inference time and training time to compare computational requirements across models.}
\label{tab:2}
\small
\begin{tabular}{lcccc}
\toprule
\textbf{Metric} & \textbf{ResNetLite} & \textbf{ResNet18} & \textbf{MammoCLIP} & \textbf{RAD-DINO} \\
\midrule
Sensitivity (Recall) & $0.9640 \pm 0.0165$ & $0.9741 \pm 0.0136$ & $0.9722 \pm 0.0143$ & $0.9802 \pm 0.0124$ \\
Specificity & $0.9880 \pm 0.0094$ & $0.9961 \pm 0.0051$ & $0.9980 \pm 0.0031$ & $0.9980 \pm 0.0031$ \\
Precision & $0.9877 \pm 0.0094$ & $0.9960 \pm 0.0052$ & $0.9980 \pm 0.0032$ & $0.9980 \pm 0.0032$ \\
F1-Score & $0.9757 \pm 0.0096$ & $0.9849 \pm 0.0079$ & $0.9849 \pm 0.0076$ & $0.9890 \pm 0.0065$ \\
Accuracy & $0.9760 \pm 0.0095$ & $0.9851 \pm 0.0075$ & $0.9851 \pm 0.0075$ & $0.9891 \pm 0.0065$ \\
AUROC & $0.9929 \pm 0.0046$ & $0.9943 \pm 0.0051$ & $0.9989 \pm 0.0013$ & $0.9986 \pm 0.0016$ \\
AUPRC & $0.9938 \pm 0.0042$ & $0.9951 \pm 0.0049$ & $0.9989 \pm 0.0014$ & $0.9989 \pm 0.0013$ \\
\midrule
Parameters & 154,978 & 11,177,538 & 28,340,784 & 86,580,480 \\
Inference time & 421.65 ms & 431.81 ms & 446.19 ms & 493.08 ms \\
Training time & 675 s & 683.3 s & 493 s & 784.4 s \\
\bottomrule
\end{tabular}
\end{table}

When examining AUROC and AUPRC, both models performed at the upper limit of discriminative ability, with MammoCLIP achieving the highest AUROC among all four models (0.9989 vs. 0.9986 for RAD-DINO); this difference was likewise not statistically significant given the overlapping confidence intervals. Importantly, MammoCLIP achieved this comparable performance with far fewer parameters (28.3M vs. 86.6M), suggesting greater computational efficiency and scalability. This was further supported by shorter training times (embedding generation and SVM training), as MammoCLIP finished training in 493 seconds compared to 784.4 seconds for RAD-DINO.

\begin{figure}[tbp]
\centering
\includegraphics[width=0.85\textwidth]{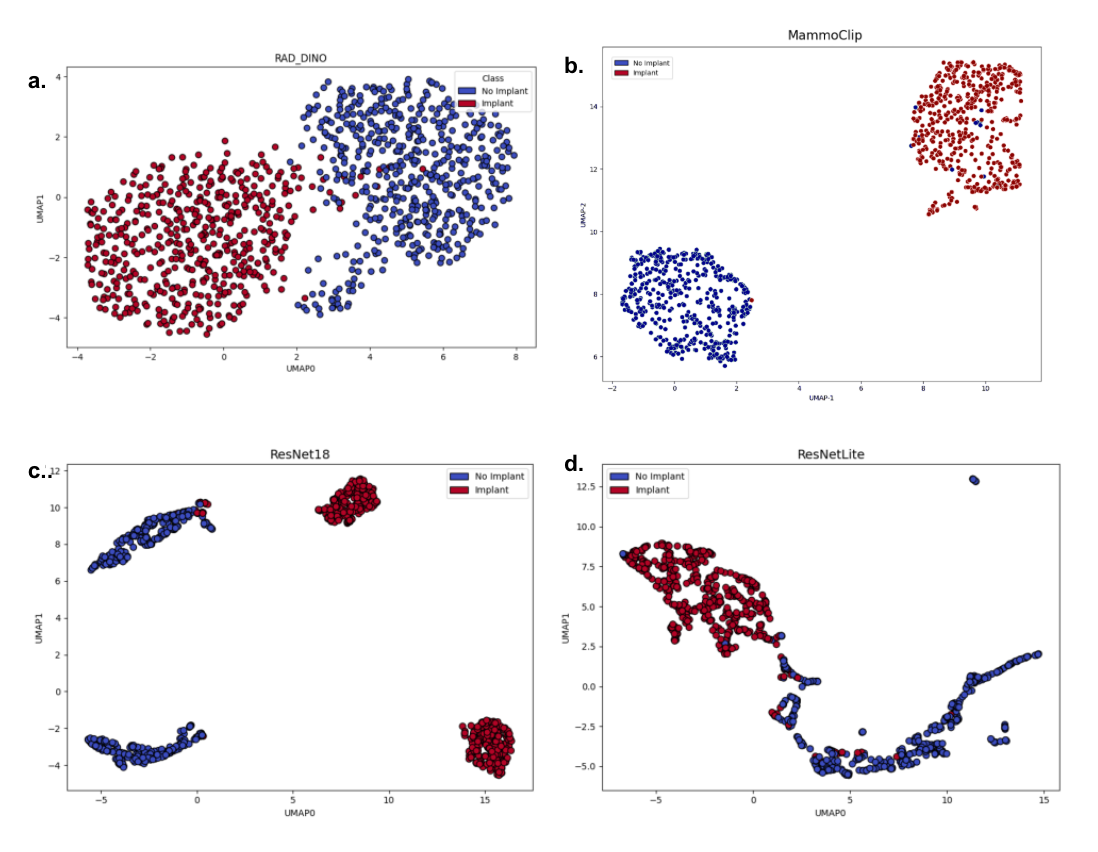}
\caption{\textbf{UMAP visualization of feature embeddings by implant status.} UMAP visualizations of feature embeddings demonstrate differences in clustering quality and separation between implant (red) and no implant (blue) cases. (a) RAD-DINO shows two distinguishable clusters, though with a degree of overlap at the boundary between classes, suggesting good but imperfect feature discrimination. (b) MammoCLIP demonstrates the tightest and most distinct separation among all models, with the two classes forming compact, well-isolated clusters and minimal overlap. (c) ResNet18 embeddings form four tight clusters, reflecting strong high-level feature representations consistent with the model's discriminative behavior before classification. (d) ResNetLite exhibits weaker clustering structure and more overlap, however, this qualitative difference in embedding structure did not correspond to a statistically significant difference in classification performance relative to ResNet18.}
\label{fig:5}
\end{figure}

The embedding spaces (Fig.~\ref{fig:5}) showed some structural differences between the two models. MammoCLIP embeddings formed two cleanly separated clusters with minimal overlap, while RAD-DINO embeddings also displayed clear separation between classes but with slightly more overlapped boundaries. Despite this qualitative difference in representation structure, it did not correspond to a statistically significant difference in classification performance.

\subsection{Comparison of Convolutional Models}

ResNet18 showed higher point estimates than ResNetLite across classification metrics (Table~\ref{tab:2}). ResNet18 achieved higher sensitivity (0.9741 vs. 0.9640) and specificity (0.9961 vs. 0.9880); however, their 95\% confidence intervals overlapped substantially for both metrics, indicating these differences are not statistically significant. Precision and F1-scores followed a similar trend, with ResNet18 showing higher point estimates (Precision: 0.9960 vs. 0.9877; F1-score: 0.9849 vs. 0.9757; Table~\ref{tab:2}), though the confidence intervals for these metrics also overlapped. Overall accuracy was likewise higher for ResNet18 (0.9851 vs. 0.9760), with overlapping confidence intervals indicating this difference is not statistically significant either.

Similarly, AUROC and AUPRC values were higher for ResNet18 (0.9943 vs. 0.9929 and 0.9951 vs. 0.9938, respectively), with overlapping confidence intervals across all metrics suggesting comparable overall discriminative ability. Despite these similar performance levels, ResNetLite utilized 154,978 parameters, representing approximately 1.4\% of the 11.2 million parameters in the ResNet18 architecture.

Despite containing approximately $72\times$ fewer parameters, ResNetLite demonstrated training times comparable to ResNet18 (675 s vs. 683 s). This suggests that training throughput was likely constrained more by data loading, image preprocessing, and GPU parallelization overhead than by parameter count alone.

The embedding spaces (Fig.~\ref{fig:5}) showed some structural differences between the two models. ResNet18 produced four tight, well-separated clusters, while ResNetLite embeddings were more diffuse, with less distinct boundaries between classes. Despite this qualitative difference in representation structure, it did not correspond to a statistically significant difference in classification performance. This suggests that ResNetLite's smaller architecture organizes discriminative information differently, but no less effectively in practice, than ResNet18's higher-capacity representation.

\subsection{Comparison across models}

Across all metrics, ResNetLite showed no statistically significant difference from ResNet18 (all $p$ $>$ 0.05), suggesting its substantially smaller architecture achieves performance parity with a full-scale CNN. Compared with the foundation models, ResNetLite showed small but statistically significant deficits relative to RAD-DINO across all metrics (sensitivity, specificity, accuracy, and AUROC; all $p$ $<$ 0.05) and relative to MammoCLIP on specificity, accuracy, and AUROC (all $p$ $<$ 0.05), though not sensitivity ($p$ = 0.332). This suggests ResNetLite's performance gap is limited to comparisons against large-scale pretrained foundation models, while remaining statistically indistinguishable from a comparably-sized task-specific CNN.

Despite architectural differences, all models demonstrated end-to-end inference times below 0.5 seconds per image, supporting the feasibility of near-real-time clinical deployment (Table~\ref{tab:2}). Among the FMs, MammoCLIP demonstrated faster embedding generation compared with RAD-DINO while maintaining comparable classification performance. Specifically, MammoCLIP generated embeddings for 5,000 images while also achieving slightly lower end-to-end inference latency compared to RAD-DINO (446 ms/image vs. 493 ms/image, respectively) ($p$ $>$ 0.05). ResNetLite achieved the fastest inference time (422 ms/image) (Table~\ref{tab:2}) compared to ResNet18 (431 ms/image). Because inference latency was comparable across architectures in a GPU environment, the cost of inference per image is expected to be similar under equivalent GPU deployment conditions. However, given the substantially lower parameter count of the CNNs, these models may be amenable to deployment on less computationally intensive hardware, including CPUs, potentially reducing infrastructure costs for hospital deployment~\cite{r31}.

\subsection{Errors}

\begin{figure}[tbp]
\centering
\includegraphics[trim=0 191bp 0 0, clip, width=0.55\textwidth]{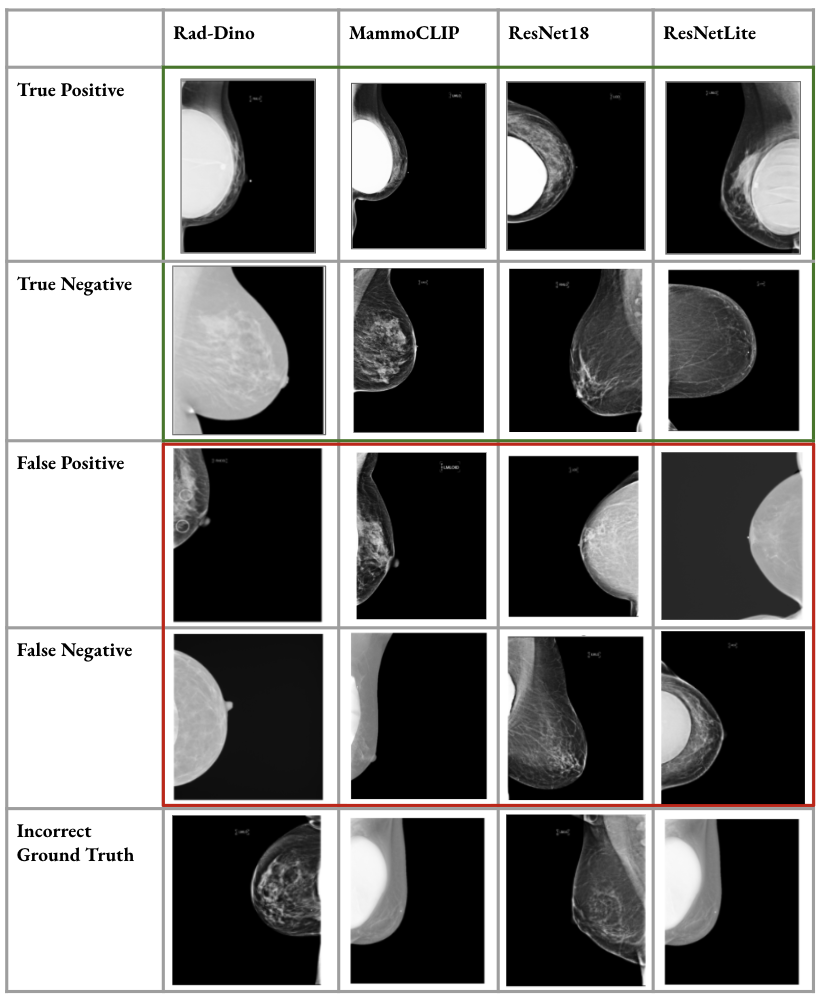}
\caption{\textbf{Examples of model predictions with classification accuracy.} Examples of model predictions are shown with green borders indicating correct classifications (true positives and true negatives) and red borders indicating incorrect classifications (false positives and false negatives). Upon visual inspection, misclassifications were more common in cases with high breast tissue density or high image contrast, suggesting that these image characteristics present challenges for the models.}
\label{fig:6}
\end{figure}

Because errors were rare and multifactorial, we performed a qualitative visual evaluation. We observed that misclassifications were most frequent in cases with extremely dense (BI-RADS density D) breast tissue across models (Fig.~\ref{fig:6})

\section{Discussion}

In this study, we evaluated the performance of two FMs and two end-to-end CNNs trained for the clinically relevant task of breast implant detection in mammographic images. Both structured clinical data and DICOM metadata are unreliable for this task; therefore, image-based classifiers provide an opportunity to detect implants directly within images, information that is crucial for downstream data curation, model development, and deployment workflows.

Our findings demonstrate that all four models achieved strong, clinically viable performance for breast implant detection, with no model showing a clear across-the-board advantage once statistical testing was applied. This is a task well-suited to real-time deployment in hospital settings, both for upstream dataset curation and as a preprocessing step ahead of downstream cancer-detection pipelines. Notably, ResNetLite showed no statistically significant difference from ResNet18 on any metric, despite using only 1.4\% of its parameters, and its gap relative to the foundation models was small and, in some cases (sensitivity vs. MammoCLIP), not statistically significant either. This has a direct practical implication: hospitals without GPU infrastructure can deploy ResNetLite or ResNet18 on minimal, low-cost hardware without a meaningful loss in diagnostic reliability, while hospitals with GPU resources available can use the foundation models, particularly MammoCLIP, which offers the highest AUROC among all four models at a fraction of RAD-DINO's parameter count. Rather than one model being the preferred choice, model selection can reasonably be guided by local infrastructure constraints.

ResNetLite illustrates this tradeoff well. Its small footprint (154,978 parameters, roughly 1.4\% of ResNet18's) allows deployment on CPU-only or resource-constrained systems, and paired bootstrap testing confirmed it performs comparably to ResNet18 across every metric tested. Its modest, statistically significant gap relative to the foundation models suggests architectural scale still confers some benefit, but that benefit is small enough that hardware availability, not accuracy, is likely to be the deciding factor in most deployment settings. This aligns with emerging evidence that architectural design, rather than model size alone, is a critical determinant of accuracy and robustness in radiology tasks~\cite{r32}.

Misclassifications occurred most frequently in dense breasts suggesting that implant detection performance may still be affected by challenging imaging contexts~\cite{r33}. Larger architectures demonstrated slightly greater robustness to these sources of variation, highlighting a potential tradeoff between efficiency and generalizability.

Finally, developing robust image-based implant detectors has thus far remained a challenge due to lack of high-quality labeled datasets. Our own labeling attempts highlighted this difficulty: relying purely on metadata-derived labels produced a 23\% error rate in identifying exams with implants, showing that this method would be both unreliable and unusable at inference time and highlighting the need for image based classification methods.

This study has three main limitations. First, we evaluated models only in the context of screening mammography, excluding diagnostic exams. This could restrict the generalizability of our findings to broader clinical workflows. However, given that FM embeddings performed very well out of the box on screening mammograms, it is reasonable to expect similar performance in diagnostic cases. Second, while we highlight the potential of smaller models, we did not perform extensive architectural comparisons beyond ResNet-based designs, leaving open the question of whether other lightweight backbones may further improve accuracy or robustness~\cite{r34}. Lastly, this is a single institution study using a single dataset. While we expect similar trends across external datasets, our trained CNNs may not directly generalize to external data without fine-tuning.

\section{Conclusion}

In conclusion, all four architectures achieved strong breast implant detection performance on held-out screening mammograms, with accuracy ranging from 0.976 (ResNetLite) to 0.989 (RAD-DINO). In GPU-equipped settings, MammoCLIP offers the highest AUROC at one-third the parameter count of RAD-DINO, making it well suited for scalable hospital deployment. In settings without dedicated GPU hardware, ResNetLite and ResNet18 would provide comparable performance at a fraction of the model size, supporting deployment on CPU-only or edge devices. These results demonstrate that the choice of model for clinical breast implant detection can be guided by local infrastructure constraints rather than performance differences alone, and support the feasibility of near-real-time implant detection for both dataset curation and downstream clinical AI workflows.

\end{document}